\documentclass[useAMS,usenatbib]{mn2e}

\usepackage{natbib}
\usepackage{graphicx}
\usepackage{savesym}
\usepackage{listings}
\usepackage[intlimits]{amsmath}
\savesymbol{iint}
\usepackage{txfonts}
\usepackage{bm}
\restoresymbol{TXF}{iint}
\usepackage{amssymb}
\usepackage{hyperref}
\usepackage{amsmath}
\usepackage{tabularx} 

\def\be{\begin{equation}}
\def\ee{\end{equation}}
\def\ba{\begin{eqnarray}}
\def\ea{\end{eqnarray}}
\def\go{\mathrel{\raise.3ex\hbox{$>$}\mkern-14mu
             \lower0.6ex\hbox{$\sim$}}}
\def\lo{\mathrel{\raise.3ex\hbox{$<$}\mkern-14mu
             \lower0.6ex\hbox{$\sim$}}}

\title[Magnetar Activity via the Density-Shear Instability]{Magnetar Activity via the Density-Shear Instability in Hall-MHD}

\author[K.N. Gourgouliatos et al.]{{Konstantinos N. Gourgouliatos\thanks{Email: K.N.Gourgouliatos@leeds.ac.uk}$^{1}$, Todor Kondic$^{1}$, Maxim Lyutikov$^{2}$ \& Rainer Hollerbach$^{1}$}\vspace{0.4cm}\\
\parbox{\textwidth}{$^{1}$Department of Applied Mathematics, University of Leeds, Leeds LS2 9JT , UK,\\
 $^{2}$Department of Physics and Astronomy, Purdue University, 525 Northwestern Ave, West Lafayette IN, 47906, USA} }

\begin{document}

\date{Accepted -. Received -; in original form -}
\pagerange{\pageref{firstpage}--\pageref{lastpage}} \pubyear{-}
\maketitle

\label{firstpage}

\begin{abstract}
We investigate the density-shear instability in Hall-MHD via numerical simulation of the full non-linear problem, in the context of magnetar activity. We confirm the development of the instability of a plane-parallel magnetic field with an appropriate intensity and electron density profile, in accordance with analytic theory. We find that the instability also appears for a monotonically decreasing electron number density and magnetic field, a plane-parallel analogue of an azimuthal or meridional magnetic field in the crust of a magnetar. The growth rate of the instability depends on the Hall properties of the field (magnetic field intensity, electron number density and the corresponding scale-heights), while being insensitive to weak resistivity. Since the Hall effect is the driving process for the evolution of the crustal magnetic field of magnetars, we argue that this instability is critical for systems containing strong meridional or azimuthal fields. We find that this process mediates the formation of localised structures with much stronger magnetic field than the average, which can lead to magnetar activity and accelerate the dissipation of the field and consequently the production of Ohmic heating. Assuming a $5\times10^{14}$G magnetic field at the base of crust, we anticipate that magnetic field as strong as $10^{15}$G will easily develop in regions of typical size of a few $10^{2}$ meters, containing magnetic energy of $10^{43}$erg, sufficient to power magnetar bursts. These active regions are more likely to appear in the magnetic equator where the tangential magnetic field is stronger. 
\end{abstract}


\section{Introduction}

The magnetic field evolution in the crust of neutron stars (NSs), in the magnetar regime, is mediated primarily by the Hall effect and Ohmic dissipation \citep{Jones:1988, Goldreich:1992}. While the familiar picture of the Hall effect is that of the creation of a voltage across an electrical conductor, when a magnetic field is administered perpendicular to the current \citep{Hall:1880}, NS applications require that the feedback of the electric current onto the magnetic field is accounted for, leading to the realms of the Hall-Magnetohydrodynamics (Hall-MHD) description. 

As Hall-MHD provides a kinematic description, which does not correspond to an energy minimisation principle \citep{Lyutikov:2013}, there is an ongoing debate regarding the stability of magnetic configurations, turbulent cascade and the overall evolutionary behaviour of a magnetic field in this context. Motivated by the mathematical similarity of the Hall-MHD equations with the vorticity equation in fluid dynamics, which is known to initiate turbulent cascade, it has been argued \citep{Goldreich:1992, Biskamp:1996, Cho:2009} that a magnetic field should undergo Hall-induced turbulence. Plane parallel and 3-D cartesian box simulations \citep{Wareing:2009b, Wareing:2010} demonstrated that while the magnetic field adopts a characteristic power spectrum once it evolves under Hall-MHD, its temporal evolution in real space consists of frozen-in structures, whose time average is non-zero unlike normal turbulence. This result is in line with the consensus of axially-symmetric spherical-shell simulations of the magnetic field evolution in NS crusts where the Hall effect operates, with subdominant Ohmic dissipation. These simulations \citep{Hollerbach:2002, Hollerbach:2004, Pons:2009, Kojima:2012, Vigano:2013, Gourgouliatos:2014b,Gourgouliatos:2014a, Marchant:2014} find that while the magnetic field may change drastically compared to its initial state as a result of the Hall effect, the evolution saturates in a short time and the system relaxes to a particular spatial structure, a result recently confirmed through 3-D spherical shell simulations \citep{Wood:2015}.  

In parallel to the turbulent cascade discussion, the question of Hall instability has been addressed. Although the Hall effect conserves magnetic energy, it can potentially drive instability by transferring energy from a Hall equilibrium state to a weaker perturbing field. Numerical and analytical studies have explored Hall instability \citep{Rheinhardt:2002, Rheinhardt:2004, Pons:2010} in cartesian geometry. Quite remarkably, axially symmetric simulations in spherical shells did not find any evidence for the operation of Hall instability, i.e.~when a state of Hall equilibrium \citep{Gourgouliatos:2013a} is chosen as an initial condition on an axially symmetric Hall simulation the system evolves because of Ohmic decay rather than the Hall effect \citep{Marchant:2014}.  

\cite{Wood:2014} studied analytically the density-shear instability for a unidirectional magnetic field. In this instability it is critical that both the magnetic field and the electron number density have strong gradients in the direction normal to the magnetic field. NS crusts are excellent environments for this instability to operate. They have a thickness of $\sim 1$km and host magnetic fields that could reach strengths of $\sim 10^{15}$G for magnetars. The density at the base of the crust approaches the nuclear density $\sim 10^{14}$g~cm$^{-3}$, and the Hall effect operates down to $\sim 10^{10}$g~cm$^{-3}$, as below this value, the effect of Lorentz forces becomes comparable to the breaking strain of the crust invalidating the Hall approach \citep{Gourgouliatos:2015}. As the chemical composition of the crust changes with depth and consequently the electron number fraction, the electron number density in which the magnetic field evolved because of Hall-MHD ranges between $\sim10^{36}$cm$^{-3}$ and $10^{34}$cm$^{-3}$ \citep{Cumming:2004}. 

Transient activity of magnetars, in the form of bursts has been attributed to Hall evolution \citep{Thompson:2001}, via crust yielding. Elaborating on this scenario, \cite{Perna:2011} used axially symmetric Hall simulations to compare the magnetic stresses exerted on the crust to the breaking strain. They found that for initial poloidal fields $B_{p}=8\times 10^{14}$G and toroidal $B_{t}=2\times 10^{15}$G magnetar activity is feasible, however a weaker initial magnetic field combination ($B_{p}=2\times 10^{14}$G and $B_{t}=10^{15}$G) leads only to sporadic bursts. Given that a substantial fraction of magnetars, have poloidal magnetic fields well below $5\times 10^{14}$G \citep{Olausen:2014}, it puts in question the validity of this scenario, given that magnetar behaviour has been observed by NSs with modest spin-down inferred dipole magnetic fields \citep{Gavriil:2002, Rea:2010, Scholz:2012}. A possible solution to this puzzle is the presence of localised stronger magnetic fields compared to the large scale ones, a scenario that has been supported observationally \citep{Tiengo:2013}. In this work we show that the density-shear instability can severely deform the large scale structure of the magnetic field in the crust of a NS and increase its intensity in areas of characteristic length-scale of a few $10^{2}$m, concentrating $10^{43}$erg of magnetic energy in the corresponding volume. This stronger magnetic field exerts stresses in the crust that can lead to yielding and eventually to magnetar bursts.   

The plan of the paper is as follows: In Section 2 we provide the mathematical formulation of the problem. In Section 3 we present the numerical scheme and the initial conditions chosen. We discuss the results in Section 4. We consider applications to magnetar activity in Section 5. We conclude in Section 6.

\section{Mathematical Formulation}

In the electron-MHD limit of the Hall effect, the electron fluid velocity $\bm{v}_{\rm e}$ is related to the electric current density by $\bm{j} = -{\rm e}n_{\rm e} \bm{v}_{\rm e}$, where $n_{\rm e}$ is the electron number density, $c$ and ${\rm e}$ are the speed of light and the electron elementary charge. Then, from Amp\`ere's law, the electric current density is $\bm{j}=\frac{c}{4 \pi}\nabla \times \bm{B}$, where $\bm{B}$ is the magnetic induction. We can safely neglect Maxwell's correction as the velocities involved are non-relativistic. The electron velocity becomes $\bm{v}_{\rm e} = -\frac{c}{4 \pi {\rm e} n_{\rm e}}\nabla \times \bm{B}$. Assuming some finite conductivity $\sigma$, the electric field reads $\bm{E} =-\bm{v}_{\rm e} \times \bm{B}/c+ \bm{j}/\sigma$. Finally we substitute into Faraday's law to obtain the induction equation:
\begin{eqnarray}
\frac{\partial \bm{B}}{\partial t} = -\nabla \times \left(\frac{c}{{4 \pi \rm e}n_{\rm e}} \left(\nabla \times \bm{B}\right)\times \bm{B} +\frac{c^{2}}{4 \pi \sigma} \nabla \times \bm{B}\right)\,.
\label{HALL}
\end{eqnarray} 
The first term in the right hand side of equation ~(\ref{HALL}) describes the evolution of the magnetic field under the influence of the Hall effect, while the second term describes Ohmic dissipation. We define a timescale for the Hall effect $t_{H} = \frac{4 \pi {\rm  e}n_{\rm e}L^{2}}{c|B|}$, where $L$ is the typical length scale of the problem, while for Ohmic decay it is $t_{O}=\frac{4 \pi \sigma L^{2}}{c^{2}}$; the ratio of $t_{O}/t_{H}$ gives the dimensionless Magnetic Reynolds number $R_{B}=\frac{\sigma |B|}{c {\rm e}n_{\rm e}}$, also referred to as the Hall Parameter. 

Having assumed a plane-parallel geometry, the system is invariant to translations in the $y$ direction and the quantities depend only on $x$ and $z$. We then express the magnetic field in terms of two scalar functions: 
\begin{eqnarray}
\bm{B} =\nabla \Psi(x,z) \times \hat{\bm{y}} + B_{y}(x,z) \hat{\bm{y}}\,, 
\label{BFIELD}
\end{eqnarray}
which is by construction divergence free. Substituting expression (\ref{BFIELD}) into the induction equation (\ref{HALL}), we obtain two coupled differential equations for $B_{y}$ and $\Psi$:   
\begin{flalign}
\frac{\partial \Psi}{\partial t} &= \frac{c}{4 \pi n_{\rm e} {\rm e}} \left(\nabla B_{y} \times\hat{\bm{y}} \right) \cdot \nabla \Psi + \frac{c^{2}}{4 \pi \sigma } \nabla^{2} \Psi\,, \label{PSI}\\
\frac{\partial B_{y}}{\partial t} &= -\frac{c}{4 \pi {\rm e}}\left[ \left(\nabla \left(\frac{\nabla^{2}\Psi}{n_{\rm e}}\right)\times  \hat{\bm{y}}\right)\cdot \nabla \Psi + B_{y}\left(\nabla n_{\rm e}^{-1} \times \hat{\bm{y}}\right)\cdot \nabla B_{y}\right] \nonumber \\
&+\frac{c^{2}}{4 \pi \sigma}\left(\nabla^{2} B_{y} -\sigma^{-1} \nabla B_{y} \cdot \nabla \sigma\right)\,.
\label{BY}
\end{flalign} 
We switch to dimensionless quantities, keeping the same notation.
\begin{flalign}
\frac{\partial \Psi}{\partial t} &= n_{\rm e}^{-1}\left(\nabla B_{y} \times\hat{\bm{y}} \right) \cdot \nabla \Psi + R_{B}^{-1} \nabla^{2} \Psi\,, \label{NORMPSI} \\
\frac{\partial B_{y}}{\partial t} &= -\left[ \left(\nabla \left(\frac{\nabla^{2}\Psi}{n_{\rm e}}\right)\times  \hat{\bm{y}}\right)\cdot \nabla \Psi + B_{y}\left(\nabla n_{\rm e}^{-1} \times \hat{\bm{y}}\right)\cdot \nabla B_{y}\right] \nonumber \\
&+R_{B}^{-1}\left(\nabla^{2} B_{y} -\sigma^{-1} \nabla B_{y} \cdot \nabla \sigma\right) \,.
\label{NORMBY}
\end{flalign} 
Appropriate profiles of $n_{\rm e}(x)$ are imposed; $\sigma$ is taken to be a constant. In our simulation the unit time is $t_{H}$, while the Ohmic dissipation time is $R_{B}t_{H}$. Because of the varying electron density and magnetic field throughout the domain, Hall evolution may develop substantially faster than this timescale.

\section{Numerical Simulation}

We integrate the full non-linear equations (\ref{NORMPSI}) and (\ref{NORMBY}), using Euler's method, in a uniform grid $i,k$ so that $x=idx$ and $z=kdz$. We apply periodic boundary conditions in $z$. Regarding the $x$ boundary condition we use two setups. First, we assume a vacuum in either side of the $x$ boundary, by fitting a current-free magnetic field for $x<-1$ and $x>1$ (BC1); in this setup the large scale magnetic field and the electron density profile are symmetric about the axis $x=0$. This is used to confirm the occurrence of the instability and to compare with the analytical model. In the second setup, we use the vacuum boundary condition for $x>1$, while for $x<-1$ we assume that there is no magnetic field penetrating that boundary, by setting $\Psi=B_{y}=0$ at $x=-1$ (BC2). This condition is more restrictive than the Meissner superconductor boundary condition regarding the $B_{y}$ component \citep{Hollerbach:2004}, nevertheless, it is a good approximation once $R_{B}\gg 1$, as is the case here. In this configuration, the electron number density has its maximum values at $x=-1$ and decrease monotonically, resembling the structure of a NS crust. We implement these boundary conditions by using an appropriate set of ghost points. We have tested the results in different resolution levels to ensure their validity. We use a courant condition that adjusts the timestep depending on the maximum electron velocity.

It has been shown analytically that the density-shear instability occurs when the (dimensionless) magnetic field and the electron number density profiles are chosen so that $B_{z}(x)=n_{\rm e}(x)={\rm sech}^{\gamma}(x)$, where $\gamma$ is some positive constant \citep{Wood:2014}. Taking the asymptotic limit for $\gamma \to 0$ for the expression ${\rm sech}^{\gamma}(x/\gamma)$ and $\gamma \to \infty$ of the expression ${\rm sech}^{\gamma}(x/\gamma^{1/2})$ we find respectively the backgrounds $B=n_{\rm e}=\exp(-|x|)$ and $B=n=\exp(-x^2/2)$. In this work we focus in the gaussian profile because of its smoothness. We have also run some simulations  using the absolute value profile to validate the occurrence of the instability. 

We implement these profiles as follows. The absolute value profile where the initial condition for the magnetic field is $\bm{B}= B_{0}\left(\exp(-|x/L_{B}|) +\epsilon_{B}\right)\hat{\bm{z}}$ and $n_{\rm e}=n_{0}\left(\exp(-|x/L_{n}|)+\epsilon_{n}\right)$, and the gaussian profile with initial magnetic field $\bm{B}=B_{0}\left(2\pi^{-1/2} \exp(-x^{2}/L_{B}^{2}) +  \epsilon_{B}\right)\hat{\bm{z}}$ and $n_{\rm e}=n_{0}\left(\exp(-x^{2}/L_{n}^{2})+\epsilon_{n}\right)$. We superimpose a perturbation term $\bm{b}=-\delta b (\cos(k_p z) \bm{\hat{x}} + \sin(k_p z) \bm{\hat{y}})$. We have included a uniform background field $\epsilon_{B}B_{0}\hat{\bm{z}}$ and a uniform background density $\epsilon_{n}n_{0}$, with $\epsilon_{n} \ll 1$ and $\delta b \ll \epsilon_{B}B_{0}$, to ensure that the perturbing magnetic field and currents are always subdominant compared to the background field; the typical values used for the background field is $10^{-2}B_{0}$ and the perturbation $10^{-4}B_{0}$. The above profiles are used with the boundary condition BC1 where the system is symmetric with respect to $x=0$. We also used a translated version of the gaussian profile where $x\to x+1$, imposing BC2,  in this case the code dissipates some energy to force the perturbation on $B_{y}$ to satisfy the boundary condition at $x=-1$. 

We also run simulations using a pseudo-spectral parallel code. This code implements the second order, 
Runge-Kutta ETD time-stepping scheme described in \cite{Cox:2002}, and has been modified to integrate the Hall-MHD equations. The main difference is that we employ periodic boundary conditions both in x and z boundaries, unlike the grid-based one which assumes vacuum or the non-penetrating field condition in the x direction.

\section{Results}

We have explored various combinations of the parameters. A summary is shown in Table 1, where we provide information on the initial conditions, and the resulting instability. In addition to the quantities already defined we give the wavenumber of the fastest growing mode $k_{i}$, the corresponding growth timescale $\tau$ and the resolution used. 

We have confirmed that a uniform magnetic field on a gaussian density background (and vice versa, runs S1 and S2) does not lead to any unstable mode. We have run simulations using the gaussian profile and BC1 boundary conditions for a broad combination of parameters (G), we considered no background uniform field $\epsilon_{B}=0$, while keeping the other quantities the same (G9) which also gave rise to the instability. We have also used a smaller number of simulations using the absolute value profile (A) and BC1 boundary conditions. Applying BC2, we run two simulations (C) using the translated gaussian profile $\bm{B}=B_{0}\left(2\pi^{-1/2} \exp(-(x+1)^{2}/L_{B}^{2}) +  \epsilon_{B}\right)\hat{\bm{z}}$ and $n_{\rm e}=n_{0}\left(\exp(-(x+1)^{2}/L_{n}^{2})+\epsilon_{n}\right)$. 

\begin{figure}
\includegraphics[width=\columnwidth]{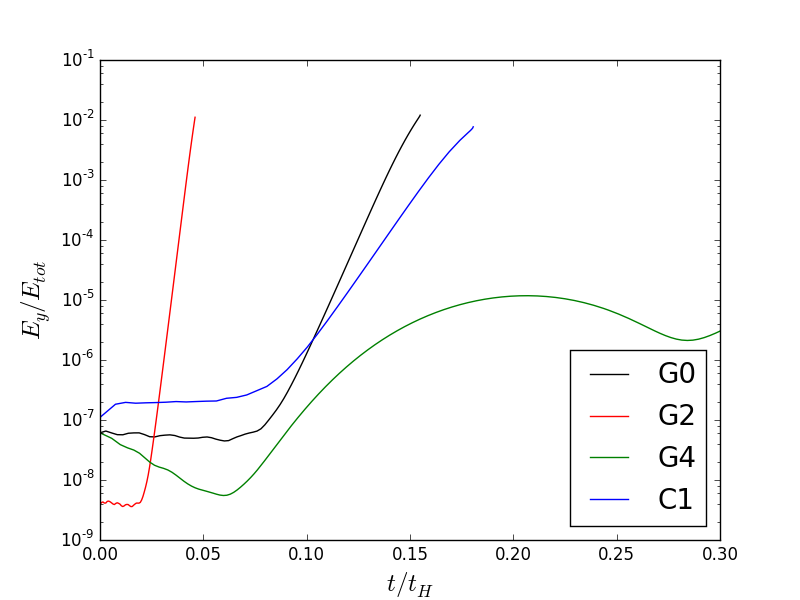}
\caption{The ratio of the energy in the $B_{y}$ component over the total magnetic energy for some characteristic runs (for details on the parameters refer to Table 1). With the exception of the highly dissipative model G4, all other models undergo some rearrangement of the perturbing field which is is followed by exponential growth of the instability.}
\label{FIG:1}
\end{figure}

\begin{figure}
\includegraphics[width=1.0\columnwidth]{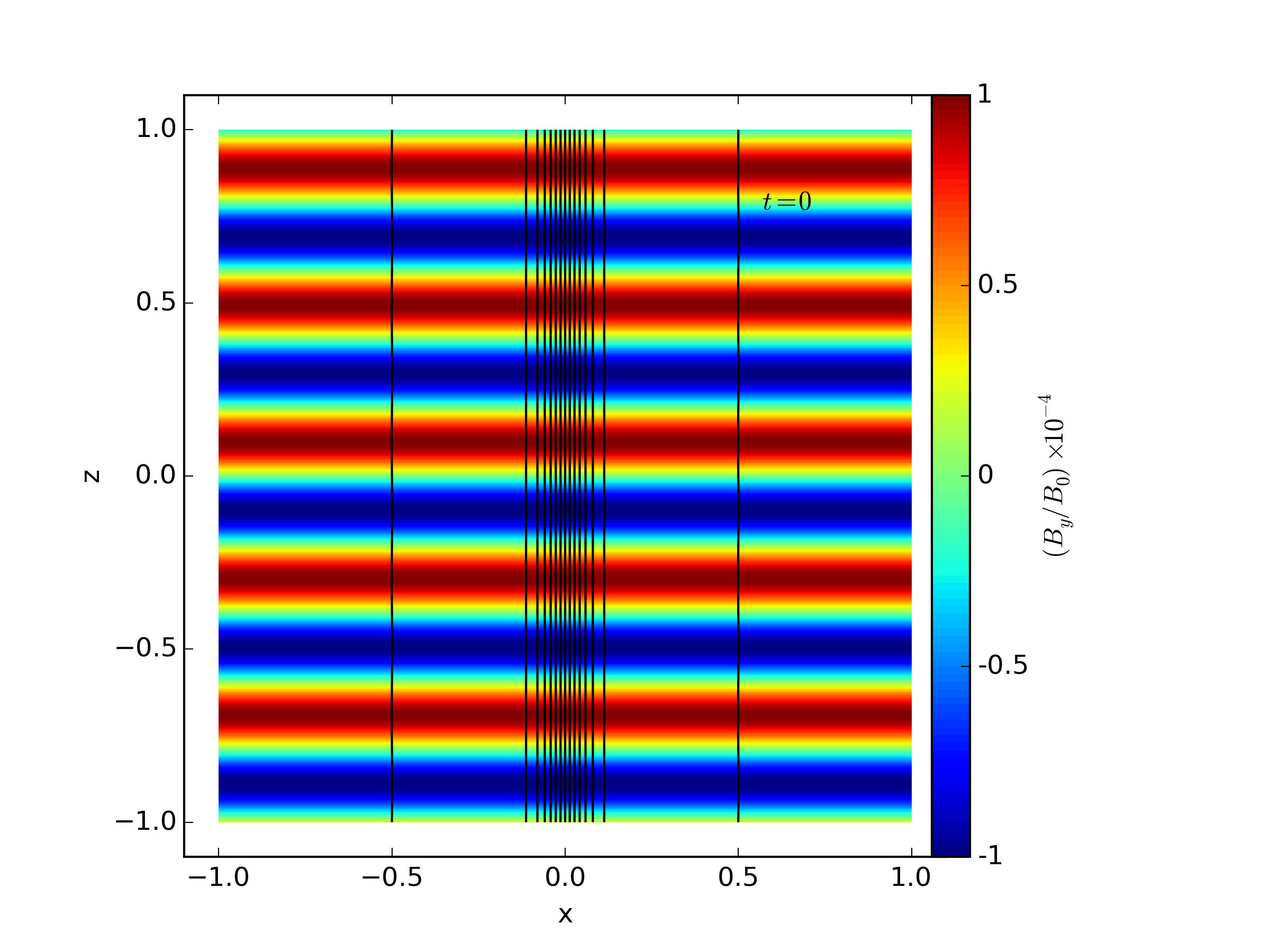}
\includegraphics[width=1.0\columnwidth]{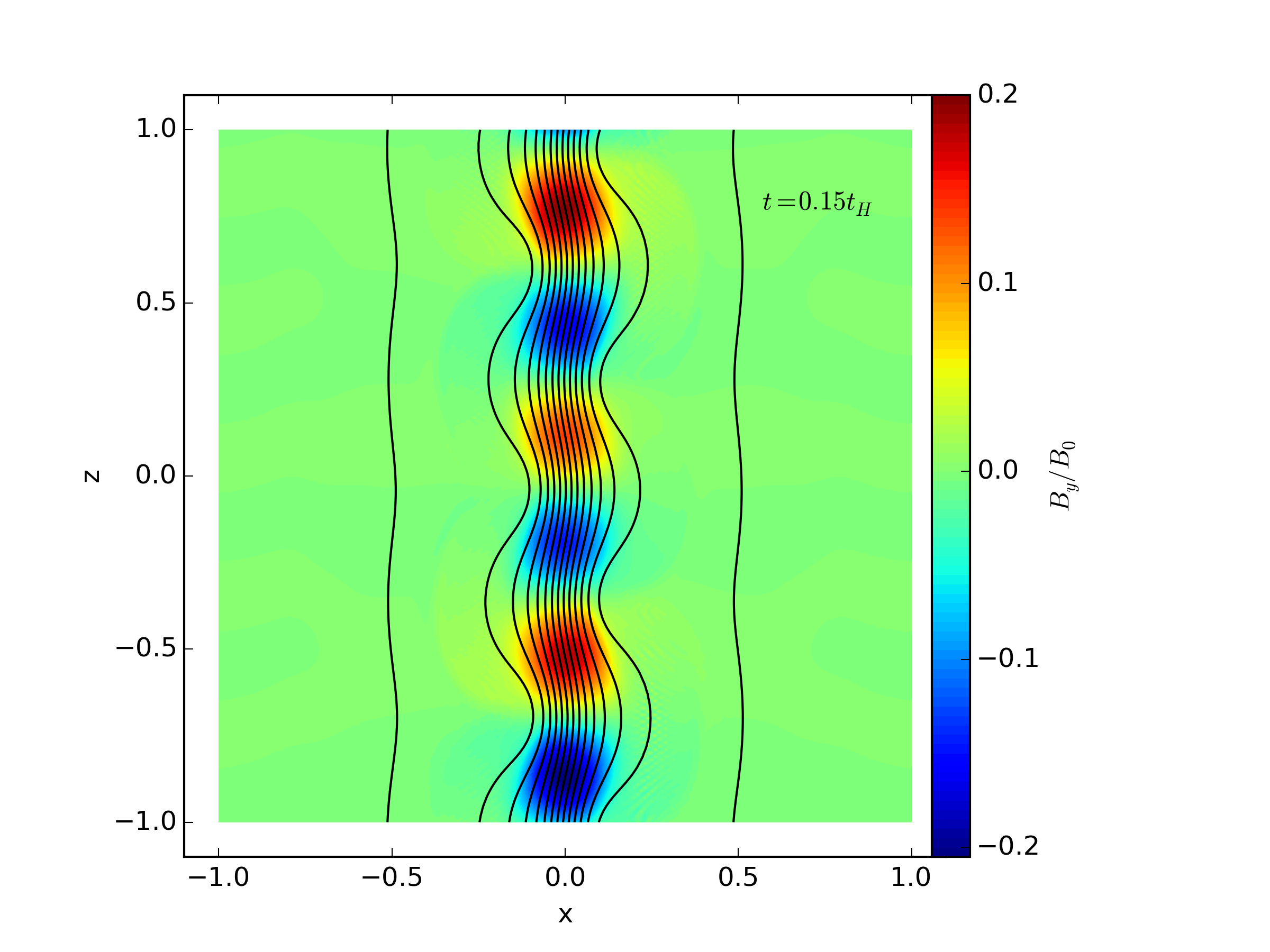}
\caption{The structure of the magnetic field at the beginning of the simulation (top panel) and once the instability has fully developed at $t=0.15t_{H}$ (bottom panel), for the simulation G0. The $B_{x}$ and $B_{z}$ components are plotted in black, while the $B_{y}$ component is shown in colour.}
\label{FIG:2}
\end{figure}

We confirm the development of the instability once the scale-height of the magnetic field and the density variation are comparable within a range of a few, and the resistivity is weak. The system undergoes some adjustment, followed by exponential growth of the instability, see Fig.~\ref{FIG:1}. Once the instability fully develops, its energy content is comparable to that of the background magnetic field, with the overall structure being deformed, Fig.~\ref{FIG:2}. At this point  large electron velocities develop, leading to a very small timestep forcing us to stop our calculation, a numerical limitation known to exist in explicit Eulerian Hall-MHD simulations \citep{Falle:2003}. The wavenumber of the fastest growing mode of the instability depends on the scale height of the magnetic field and electron number density, being inversely proportional to them once $L_{B}=L_{n}$, see for instance the $k_i$'s of G0 and G7. However if the scale-heights of the magnetic field and the density are not equal, the evolution becomes more complex, with the magnetic field needing extra time to adjust to the density background before the instability starts growing (G0 vs G6). The growth rate is proportional to the strength of the magnetic field, i.e.~G0, G1 and G2, where the ratio of the respective $\tau$'s is 0.5 while $B_{0}$ is increasing by 2. In the limit of strong resistivity (i.e.~G4), the instability may be suppressed, without dominating the overall evolution, even though there is some modest growth at the beginning. 

Repeating the analysis of \cite{Wood:2014}, for the gaussian profile under our normalisation for $L_{B}=L_{n}=L$, we find that the growth rate $\omega^{2}=B_{0}^{2}k_{i}^2(2-L^{2}k_{i}^{2})/(L^{2}\pi n_{0})$, with the maximum rate occurring for $k_{i}=L^{-1}$ giving $\omega_{max}=B_{0}/(L^{2}\sqrt{\pi n_{0}})$. As the wavenumber of the fastest growing mode of the instability is small, it is affected by the size of the simulation box, being forced to be a multiple of $\pi$ because of the periodic boundary conditions imposed. Having assumed $L_{n}=L_{B}=0.1$, $n_{0}=1=B_{0}$ (G8), the simulation gives $k_{i}=3\pi=9.42$ versus an analytical value of $k_{i}=0.1^{-1}=10$ and a corresponding growth timescale $\tau(=\omega^{-1})= 0.0138$ versus an analytical prediction of 0.0178. This deviation is due to numerical constraints and also to the superimposed uniform magnetic field and background density which are not present in the analytical model. Because of numerical limitations we have not been able to set a strong constraint on the maximum and minimum ratio of $L_{B}/L_{n}$ where the instability appears, except for the fact that there is no instability for uniform magnetic field or density. To investigate that, it would require either $L_{B}\ll L_{n}$ or $L_{B}\gg L_{n}$ and both of them to be much smaller than the size of the box, leading to a calculation that ranges over a few orders of magnitude. In the simulations where we used the absolute value profile, even at a very low resistivity (A1) the system undergoes some significant decay as the currents are very strong around $x=0$, which slows down the growth of the instability, while choices of higher resistivity (A2) prevent its development entirely. 

\begin{figure}
\includegraphics[width=1.0\columnwidth]{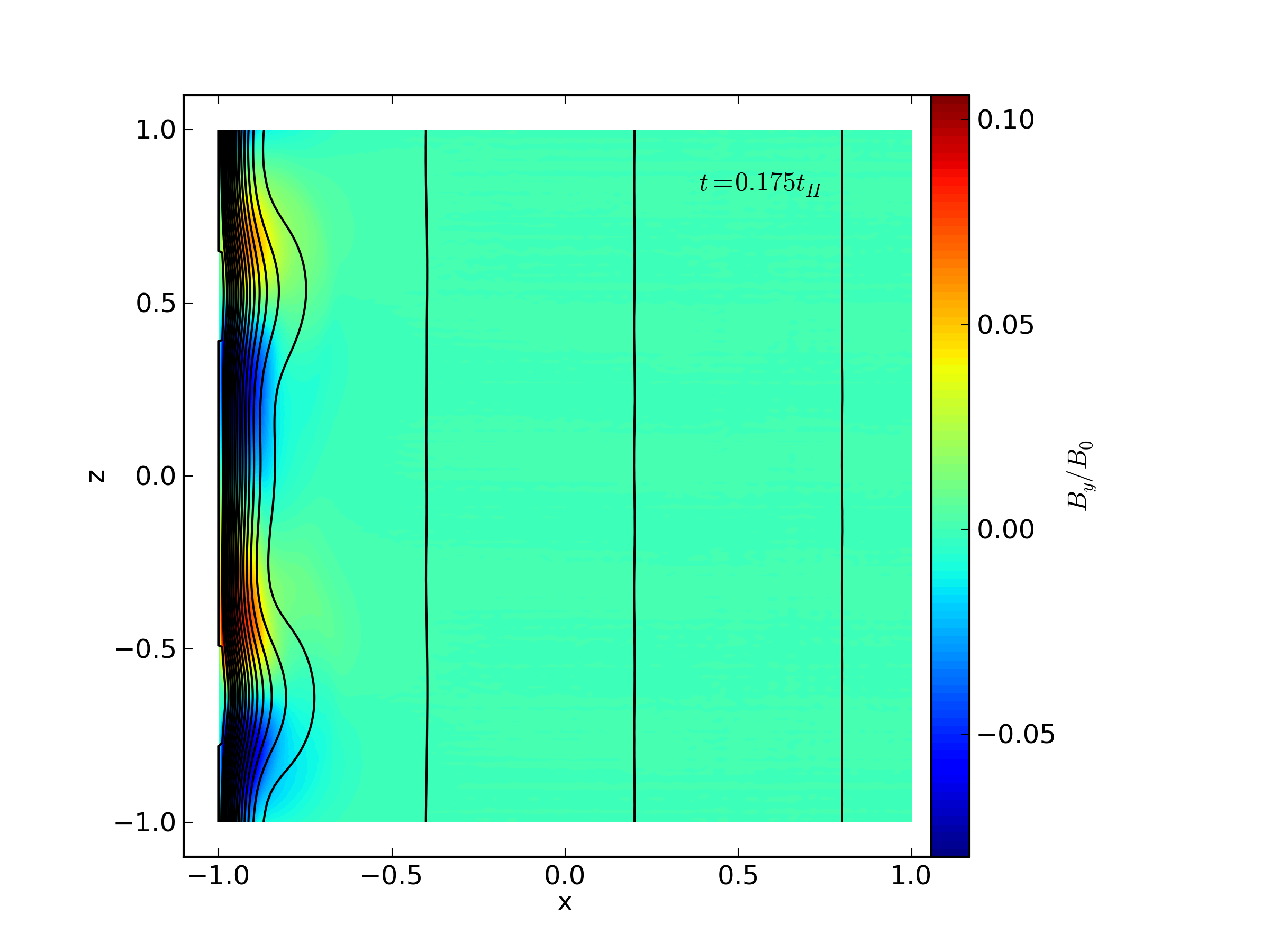}
\caption{The structure of the magnetic field once the instability has fully developed at $t=0.175$, for the simulation C1, where BC2 is used. }
\label{FIG:3}
\end{figure}

Similar behaviour is found when BC2 is applied, Fig.~\ref{FIG:3}. Given that there is a rigid boundary at $x=-1$ there is a significant growth of the $B_{x}$ and $B_{z}$ components of the magnetic field, because of the compression of the magnetic field lines against the boundary, compared to the other case, where the main effect of the instability was to kink the structure of the field. These results are in broad qualitative agreement with the linear calculation of \cite{Rheinhardt:2004}, done in a similar setup. Using the parallel code, we simulated the development of the instability for $R_{B}=50$ on a collocation  grid with $256^{2}$ points and the same initial conditions and density profiles as the G0 run (see run SP in Table 1), finding the same  behaviour.

\begin{table}
\caption{ Simulations summary. The S runs have either a uniform magnetic field or density background, the G runs utilise the gaussian profile, the A runs the absolute value profile while the C runs utilise a monotinically decreasing magnetic field and electron number density profile with BC2 boundary conditions. } 
\centering
\begin{tabular}{l l l l r r c l c} \hline\hline
NAME & $B_0$ & $L_B$ & $L_n$ &$R_B^{-1}$& $k_p/\pi$ & $k_i/\pi$ &$\tau/10^{-2}$ & Resol. \\
S1 & 1 & 0.1 & - & 0 & 10 &  - & ~~- & 200$^{2}$\\
S2 & 1 & - & 0.1 & 0 & 10 &  - & ~~- & 200$^{2}$\\
G0 & 1&0.1&0.1&0.01&5&3& 1.16&200$^{2}$\\
G1 & 2&0.1&0.1&0.01&5&3& 0.613&200$^{2}$\\
G2 & 4&0.1&0.1&0.01&5&3& 0.323&200$^{2}$\\
G3 & 1&0.1&0.1&0.05&5&2& 1.36&200$^{2}$\\
G4 & 1&0.1&0.1&0.1&5&1& ~~-&200$^{2}$\\
G5 & 1&0.1&0.05&0.001&20&4& 0.183&200$^{2}$\\
G6 & 1&0.05&0.1&0.001&10&2& 2.76&200$^{2}$\\
G7 & 1&0.05&0.05&0.001&20&6& 0.282&200$^{2}$\\
G8& 1& 0.1 &0.1 & 0 & 5 & 3 & 1.38 &200$^{2}$ \\
G9 & 1&0.1&0.1& 0 & 10 &3 & 1.36& 200$^{2}$\\
A1& 1&0.1&0.1&0.001&10&2& 3.44&100$^{2}$\\
A2 & 1&0.1&0.1&0.005&10&1& ~~-&100$^{2}$\\
C1 & 1&0.1&0.1&0&10&2& 1.76&200$^{2}$\\
C2 & 1&0.15&0.1&0&10&2& 0.731&100$^{2}$\\
SP & 1&0.1 & 0.1 &0.02&10&3&1.83&256$^{2}$\\
\hline  \hline
\end{tabular}
\label{tab:LPer}
\end{table}


\section{Application to magnetar activity}
 
Consider a magnetar, whose magnetic field and electron number density at the base of the crust are $5 \times 10^{14}$G, $10^{36}$cm$^{-3}$ respectively and crust thickness is $1$km. Using this normalisation in eqns~(\ref{NORMPSI}) and (\ref{NORMBY}), we find that the unit time of the simulation corresponds to $\sim 10^{5}$ years in the NS's life. The scale-height for the electron number density is $\sim 0.1$km, thus we expect a growth time scale of $\sim 10^{3}$ years, as we have found that $\tau \sim 10^{-2}$. Even under the conservative assumption of a magnetic field with a tiny amount of energy being in the perturbing small scale field, it is only a mater of a few thousand years for this instability to create a strong localised magnetic field, exceeding the intensity of the background field by a factor of $2$ and giving rise to magnetic fields $\sim 10^{15}$G for this setup. The size of these structures is comparable to the scale-height multiplied by $2\pi$, thus they are expected to be $\lambda_{i}\sim 0.6 $km, each one of them containing magnetic energy $10^{43}$ erg, which is sufficient to power magnetar busts. This effect is caused entirely by the redistribution of the magnetic field via the Hall effect without appealing to the generation of any extra magnetic flux. Even if a moderately strong large scale magnetic field is present ($5\times 10^{13}-10^{14}$G), this instability leads to the formation of pockets of magnetic field significantly exceeding the average value. In our simulations we found that these features typically develop near the base of the crust rather than the surface, as in our initial condition we have chosen an exponentially decreasing profile. While it is possible that such features may develop closer to the surface, this is a question to be answered conclusively by future more realistic simulations. This is particularly interesting in the context of recent observations of strong localised magnetic features such as the one observed in SGR 0418+5729 \citep{Tiengo:2013} and the 0.2-0.7 km hotspot implied by surface emission modelling in the same system \citep{Guillot:2015}.

As the components of the magnetic field which are parallel to layers of constant density are susceptible to this instability we expect the non-radial magnetic field (meridional and toroidal) to contribute the most. In a typical large-scale poloidal dipole magnetic field structure, the meridional component is stronger away from the poles, making these instabilities more likely to develop in mid-latitudes and in the equatorial region, with respect to the magnetic dipole axis. Thus we expect bursts triggered through this mechanism to provide energy away from the poles, leading to the appearance of hot spots in the form of subpulses, of the same frequency yet different phase compared to the main pulse which is likely to be associated to the magnetic pole. Recent observations show that bursts are evenly distributed in spin phase \citep{Collazzi:2015}. In any case, the complexity of heat transport within the crust \citep{Brown:2009} and the size of the active region \citep{Baubock:2015}, are critical for the observational appearance of these features. 

\section{Conclusions}

In this work we have confirmed numerically the development of the density-shear instability in a plane-parallel geometry. In particular, we have found that the instability appears when the scale height of the magnetic field and the electron number density are comparable, with the growth timescale depending on the intensity of the magnetic field, the electron number density and the relevant scale-heights. This instability also appears in a monotonically decreasing electron number density and magnetic field, a structure that encapsulates the basic characteristics of a NS crust. We conclude that the density-shear instability can lead to the formation of localised strong magnetic fields, with the typical size of these areas being a few times the scale height. Realistic NS studies need to go beyond this plane-parallel geometry, test the appearance of this instability in a 3-D calculation and investigate its evolution after it has fully developed, a task which is out the capacity of the current numerical scheme. Nevertheless, it is likely that a natural NS configuration can host an appropriate magnetic field geometry that will give rise to this instability and provide an efficient mechanism for powering magnetar activity with a weaker overall magnetic field.

\section*{Acknowledgements}

We thank Dr Toby Wood for his comments on our manuscript. KNG acknowledges a CRAQ Fellowship and Purdue University for hospitality. KNG and RH were supported by STFC Grant No. ST/K000853/1. Part of the numerical simulations were carried out on the STFC-funded DiRAC I UKMHD Science Consortia machine, hosted as part of and enabled through the ARC HPC resources and support team at the University of Leeds. 

\label{lastpage}

\bibliographystyle{mnras}
\bibliography{BibTex.bib}

\end{document}